\documentclass[aps,preprint,showpacs,superscriptaddress]{revtex4-1}

\usepackage{graphicx}
\usepackage{newtxtext,amssymb,amsmath}
\usepackage{bm,textgreek}
\usepackage{ulem}
\usepackage{comment}
\usepackage{units}
\usepackage{xfrac}

\usepackage{color}
\usepackage{xcolor}

\newcounter{defcounter}
\begin{document}

\title{	Crystal growth rates in supercooled atomic liquid mixtures}

\author{Alexander Schottelius}
\affiliation{Institut f\"ur Kernphysik, J. W. Goethe-Universit\"at, 60438 Frankfurt am Main, Germany}
\author{Francesco Mambretti}
\affiliation{Dipartimento di Fisica, Universit\`a degli Studi di Milano, 20133 Milano, Italy}
\author{Anton Kalinin}
\affiliation{GSI Helmholtzzentrum f\"ur Schwerionenforschung GmbH, 64291 Darmstadt, Germany}
\author{Bj\"orn Beyersdorff}
\affiliation{Photon Science, Deutsches Elektronen-Synchrotron (DESY), 22607 Hamburg, Germany}
\author{Andre Rothkirch}
\affiliation{Photon Science, Deutsches Elektronen-Synchrotron (DESY), 22607 Hamburg, Germany}
\author{Claudia Goy}
\affiliation{Institut f\"ur Kernphysik, J. W. Goethe-Universit\"at, 60438 Frankfurt am Main, Germany}
\author{Jan M\"uller}
\affiliation{Institut f\"ur Kernphysik, J. W. Goethe-Universit\"at, 60438 Frankfurt am Main, Germany}
\author{Nikolaos Petridis}
\affiliation{GSI Helmholtzzentrum f\"ur Schwerionenforschung GmbH, 64291 Darmstadt, Germany}
\author{Maurizio Ritzer}
\affiliation{Institut f\"ur Kernphysik, J. W. Goethe-Universit\"at, 60438 Frankfurt am Main, Germany}
\author{Florian Trinter}
\affiliation{Institut f\"ur Kernphysik, J. W. Goethe-Universit\"at, 60438 Frankfurt am Main, Germany}
\affiliation{Photon Science, Deutsches Elektronen-Synchrotron (DESY), 22607 Hamburg, Germany}
\affiliation{Molecular Physics, Fritz-Haber-Institut der Max-Planck-Gesellschaft, 14195 Berlin, Germany}
\author{Jos\'e M. Fern\'andez}
\affiliation{Laboratory of Molecular Fluid Dynamics, Instituto de Estructura de la Materia, IEM-CSIC, 28006, Madrid, Spain}
\author{Tiberio A. Ezquerra}
\affiliation{Macromolecular Physics Department, Instituto de Estructura de la Materia, IEM-CSIC, 28006, Madrid, Spain}
\author{Davide E. Galli}
\affiliation{Dipartimento di Fisica, Universit\`a degli Studi di Milano, 20133 Milano, Italy}
\author{Robert E. Grisenti}\email[]{grisenti@atom.uni-frankfurt.de}
\affiliation{Institut f\"ur Kernphysik, J. W. Goethe-Universit\"at, 60438 Frankfurt am Main, Germany}
\affiliation{GSI Helmholtzzentrum f\"ur Schwerionenforschung GmbH, 64291 Darmstadt, Germany}




\maketitle

{\bf Crystallization is a fundamental process in materials science, providing the primary route for the realization of a wide range of novel materials. Crystallization rates are considered also to be useful probes of glass-forming ability.~\cite{Tang2013,Orava2014,Orava2016}. At the microscopic level, crystallization is described by the classical crystal nucleation and growth theories~\cite{Kelton2010,Jackson2004}, yet in general solid formation is a far more complex process. Particularly the observation of apparently different crystal growth regimes in many binary liquid mixtures greatly challenges our understanding of crystallization~\cite{Spaepen1984,Stipp2009,Kerrache2008,Wang2011,Fang2013,Tang2013,Yan2015,Sun2019}. Here, we study by experiments, theory, and computer simulations the crystallization of supercooled mixtures of argon and krypton, showing that crystal growth rates in these systems can be reconciled with existing crystal growth models only by explicitly accounting for the non-ideality of the mixtures. Our results highlight the importance of thermodynamic aspects in describing the crystal growth kinetics, providing a major step towards a more sophisticated theory of crystal growth.
}

The classical crystal nucleation and growth theories describe the microscopic steps by which a solid phase spontaneously forms in the supercooled liquid at some temperature $T$ below melting. Homogeneous crystal nucleation is the process of the formation by thermal fluctuations of a small, localized nucleus of the newly ordered phase in the metastable liquid~\cite{Kelton2010}. Once the nucleus has reached its critical size, it grows at a rate that within the kinetic theory of crystal growth is given by~\cite{Jackson2004}
\begin{equation}
u(T)=fa(T)\nu(T)\exp\left(-\frac{\Delta S_{\rm m}}{R}\right)\left\{1-\exp\left[-\frac{\Delta G(T)}{RT}\right]\right\},
\label{u}
\end{equation}
where $f\leq 1$ is a geometrical factor representing the fraction of atomic collisions with the crystal surface that actually contribute to the growth, $a(T)$ is a characteristic interatomic spacing that can be identified with the lattice constant, $\nu(T)$ is the crystal addition rate at the crystal/liquid interface, $\Delta S_{\rm m}$ is the molar entropy of fusion, $R$ is the universal gas constant, and $\Delta G(T)=G^{\rm L}(T)-G^{\rm C}(T)$ is the difference in liquid (L) and crystal (C) molar Gibbs free energies. In the Wilson-Frenkel (WF) theory~\cite{Frenkel1946}, the crystal addition rate is proportional to the atomic diffusivity $D(T)$, $\nu_{\rm WF}(T)= 6D(T)/\Lambda^2(T)$, and hence exhibits the strong temperature dependence associated with an activated process. Here, $\Lambda(T)=c a(T)$ is an average atomic displacement that we assume to be proportional to $a(T)$, with $c$ being a dimensionless parameter. In the collision-limited (CL) scenario~\cite{Broughton1982}, the crystal addition rate is proportional to the average thermal velocity of the particles, $\nu_{\rm CL}(T)= \sqrt{3k_{\rm B}T/m}/\Lambda(T)$, where $k_{\rm B}$ is Boltzmann's constant and $m$ is the particle's mass, and represents the extreme case in which there is no activation barrier for ordering.

At the microscopic level, the WF and CL models can be characterized by limiting time scales associated with particle motion~\cite{Hawken2019}. The weak dependence on temperature of the particle's velocity for a Boltzmann distribution describes the ballistic motion at short times. At longer times the particle motion becomes, on average, diffusive, exhibiting an Arrhenius-like dependence on temperature. Sun {\it et al.}~\cite{Sun2018} recently found by molecular dynamics (MD) simulations that the barrier-less crystal growth kinetics in supercooled Lennard-Jones (LJ) liquids and pure metals might arise from a crystalline ground state of the atoms in the disordered liquid state adjacent to the crystal/liquid interface, effectively reducing the time required for transforming the liquid to crystal. However, many crystal growth rates are difficult to treat with these classical models, especially when key variables such as composition and particle size ratio are varied in binary systems~\cite{Spaepen1984,Kerrache2008,Stipp2009,Wang2011,Fang2013,Tang2013,Yan2015,Sun2019}. The crystallization behavior of supercooled binary alloys is of particular importance, not only because of their great technological relevance, but also for studies of glass-forming ability in few-components liquids~\cite{Orava2014,Orava2016}. Simulations do show that crystal growth rates in binary alloys can be much lower than those predicted by the CL model~\cite{Kerrache2008,Tang2013,Yan2015,Sun2019}, but the WF model also fails to describe crystal growth in these systems~\cite{Kerrache2008,Wang2011}. The current situation is that there is no comprehensive theory that is able to explain such differences in crystal growth rates arising in entirely metallic liquids~\cite{Orava2014,Orava2016}.

Here, we made a significant step forward in our fundamental understanding of crystal growth by investigating by means of experiments, theoretical modeling, and MD simulations the crystallization kinetics in supercooled mixtures of argon and krypton. Rare-gas liquids are particularly attractive model systems because MD simulations indicate that crystal growth rates in supercooled LJ liquids~\cite{Broughton1982,Burke1988} are comparable to those in pure metals~\cite{Sun2018}. Furthermore, condensed argon and krypton are miscible in the whole range of composition in both the liquid and solid phases, with a phase diagram approaching that of an ideal mixture~\cite{Heastie1959}. These features, combined with the about 8\% difference between the argon and krypton atomic radii and the absence of chemical order, make liquid mixtures of these two elements the ideal laboratory realization of the simplest atomic binary systems. However, cooling rare-gas liquids to temperatures significantly below their melting points is difficult, not to mention the subsequent probing of the rapidly evolving liquid-to-solid phase transition. Our approach was based on the generation of a microscopic laminar jet in vacuum, which offers a powerful method to investigate fast structural transformations in simple supercooled atomic and molecular liquids~\cite{Kuehnel2011,Grisenti2018}.

\begin{figure*}[h]
\includegraphics[width=1\linewidth]{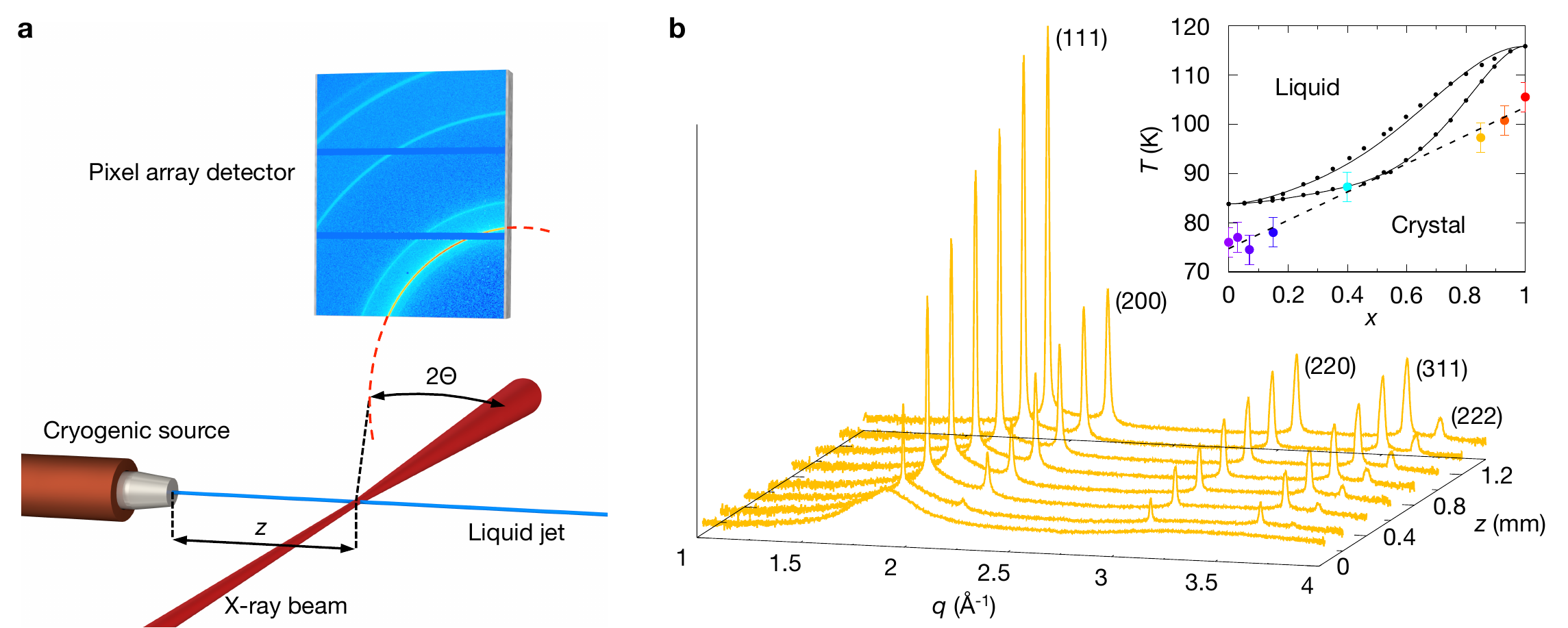}
\caption{\label{exp} {\bf Schematic of the experiment and diffraction profiles.} (a) Liquid jets with a uniform diameter of $\approx 5$~$\mu$m were generated in a vacuum and probed with a 13~keV x-ray beam. The scattered x rays were detected by a two-dimensional pixel array detector covering the 10$^{\circ}$ to 40$^{\circ}$ $2\Theta$ diffraction angular range. The scattering images resembled virtual powder diffraction patterns as a consequence of the 100~s-long acquisition time, during which up to $\sim 10^8$ individual sampled volumes crossed the $\approx 24$~$\mu$m-wide focus of the x-ray beam. (b) Selection of area-normalized diffraction profiles from a jet with $x=0.85$ as a function of the wavevector $q=4\pi \sin\left(\Theta\right) /\lambda$, where $\lambda$ is the radiation wavelength. The five most intense peaks correspond to reflections of the fcc crystal structure. The inset shows the temperature versus composition phase diagram of mixtures of argon and krypton. The solid lines are sixth-order polynomial fits to the experimental liquidus and solidus boundaries~\cite{Heastie1959}, shown as black dots. The colored filled circles represent the temperatures at the onset of crystallization in the liquid jets (see Methods), and the dashed line is a linear fit to the experimental data. The error bars result from the uncertainty in the distance between the liquid jet and the detector (see Methods).
}
\end{figure*}

Figure ~\ref{exp}a shows a schematic representation of the experiment. Liquid jets of varying krypton mole fraction $x$ between $x=0$ (pure argon jet) and $x=1$ (pure krypton jet) were generated in a vacuum chamber (see Methods). The jets cooled rapidly by surface evaporation until they crystallized spontaneously by homogeneous nucleation, forming continuous solid filaments~\cite{Grisenti2018}. We used x rays to probe the crystallization process by adjusting the distance $z$ between the orifice and the interaction region, effectively changing the time $t = z/v$, where $v$ is the jet velocity, with a resolution of $\approx 0.5$~$\mu$s (see Methods). Figure~\ref{exp}b shows reduced diffraction profiles obtained by azimuthal integration of two-dimensional scattering images from a jet with $x=0.85$ krypton mole fraction (see Methods). At small distances, the diffraction curves exhibit a broad main peak characteristic of short-range order in the disordered liquid state. With an increasing $z$, a reduction in the diffuse scattering from the liquid is accompanied by the rise in intensity of five sharp peaks, which become the dominant feature at the largest distances. We identified these peaks with reflections of the face-centered cubic (fcc) crystal structure, into which rare-gas liquids crystallize under equilibrium conditions~\cite{Pollack1964}.

\begin{figure*}[h]
\includegraphics[width=1\linewidth]{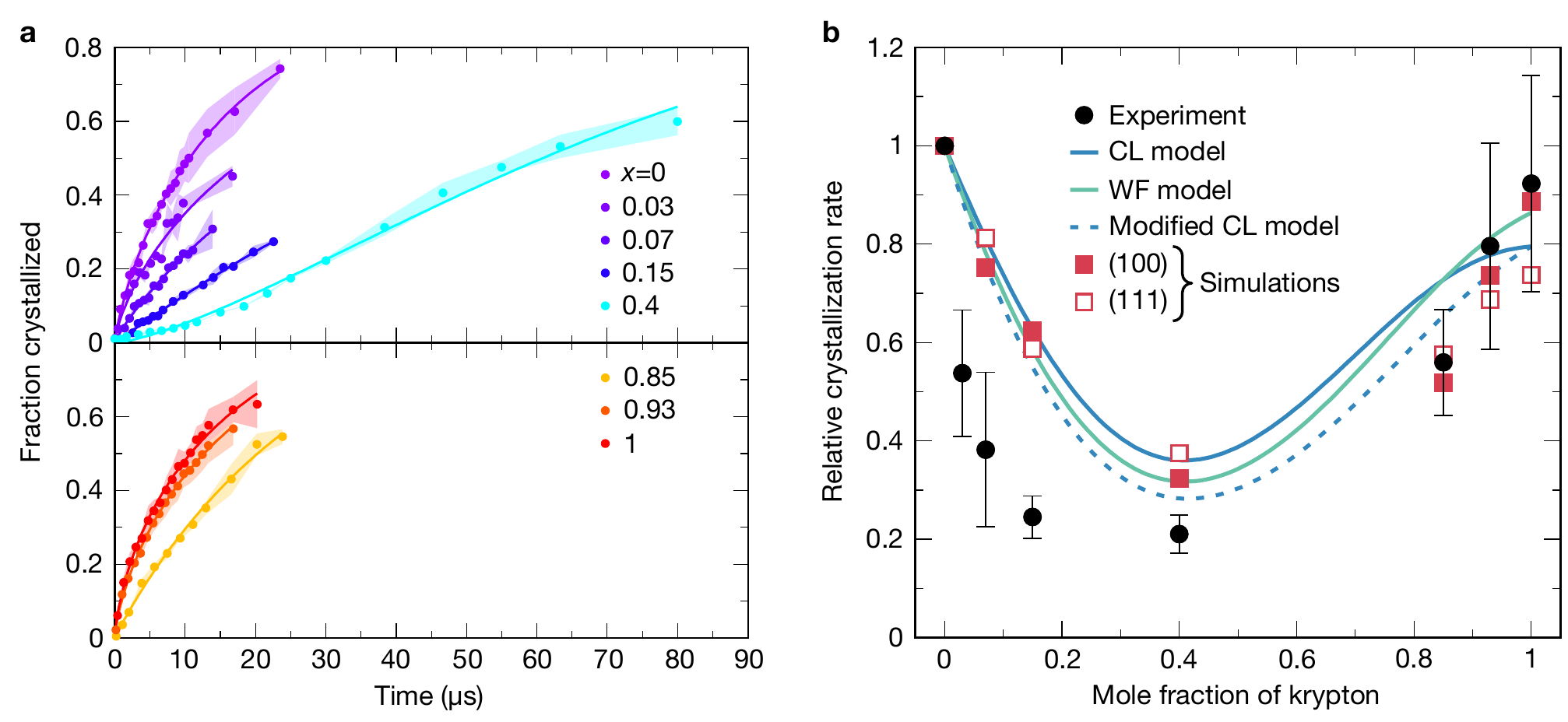}
\caption{{\bf Experimental data and comparison with theoretical calculations and simulation results.} (a) The symbols represent the time evolution of the fraction crystallized for argon- (top panel) and krypton-rich (bottom panel) mixtures obtained from the analysis of the experimental diffraction profiles. The time axis is defined as $t = z/v$. The light-shaded regions represent the estimated uncertainties in the evaluation of the fraction crystallized as described in the Methods. The solid curves are fits of the JMAK model to the experimental data. The fitted values of $\tau$ (see Supplementary Table~1) were used to shift the experimental data sets to a common origin. (b) The filled circles are the ratios of the crystallization rate constants obtained from the fits shown in (a) to the crystallization rate constant for the pure argon jet. The error bars result from the uncertainties in the experimental fraction crystallized. The blue and green solid lines are theoretical relative crystal growth rates calculated at the interpolated temperature $T(x)$ on the basis of Eq.~(\ref{u}) with crystal addition rates $\nu_{\rm CL}(T)$ and $\nu_{\rm WF}(T)$, respectively. The dotted line is the calculation with the crystal addition rate $\widetilde{\nu}_{\rm CL}(T)$ defined by Eq.~(\ref{nu_mod}). The filled and open squares are the ratios of the simulated crystal growth rates of the (100) and (111) surfaces, respectively, to the respective rates obtained for the pure argon system. The numerical parameters used for these simulations are provided in Supplementary Table~2. The error bars are smaller than the symbol size.
}
\label{X_and_u}
\end{figure*}

Figure~\ref{X_and_u}a shows the time evolution of the converted liquid fraction in the scattering volume, extracted from diffraction profiles like those shown in Fig.~\ref{exp}b (see Methods). We fitted the experimental data by the Johnson-Mehl-Avrami-Kolmogorov (JMAK) rate equation $X(t)=1-\exp\left\{-\left[k (t-\tau)\right]^n\right\}$ \cite{Jackson2004}, where $k$, $\tau$, and $n$ are fit parameters. The fits are shown as solid curves in Fig.~\ref{X_and_u}a, with the values of the fit parameters presented in Supplementary Table~1. The Avrami exponent $n$ contains information on the crystal nucleation and growth topologies. We found that $n$ is close to unity for the eight investigated jets, a result that might be associated with surface nucleation~\cite{Cahn1956}. The JMAK model provides a widely used approach to obtain a reliable estimation of the characteristic crystallization rate constant $k$. The ratio of $k$ to the crystallization rate constant for the pure argon jet is plotted in Fig. 2b as filled circles. The data indicate a slowdown of the crystallization kinetics with increasing $x$, which becomes faster again for the krypton-rich jets. In particular, the $x=0.4$ mixture crystallized, on average, at a five times smaller rate than the pure argon and krypton jets.

The trend displayed in Fig.~\ref{X_and_u}b is qualitatively similar to that reported for diverse binary systems, such as colloids~\cite{Williams2008,Stipp2009} and quantum liquids~\cite{Kuehnel2014}. However, a direct comparison with Eq.~(\ref{u}) is in general precluded because the explicit calculation of the crystal growth rate requires thermodynamic data that are either difficult to obtain or unavailable, so that several approximations must be introduced~\cite{Jackson2004}. Condensed argon and krypton and their mixtures represent a fortunate exception, as phase-equilibrium thermodynamic data are available for these systems (see Methods and Supplementary Fig.~1). We therefore carried out an exact analytic calculation of $u(T)$, the details of which are provided in Methods. Since the crystal growth rate is a function of the temperature, to allow for comparison with the experimental data we determined the jet temperature at the onset of crystallization on the basis of the lattice constants extracted from the diffraction profiles like those shown in Fig.~\ref{exp}b (see Methods and Supplementary Fig.~2). The resulting jet temperatures are plotted as colored circles in the inset of Fig.~\ref{exp}b. To a good approximation, their dependence on the krypton mole fraction can be treated as linear~\cite{Kuehnel2014}, as shown by the fit $T(x)=74.7+28.7x$ (dashed line). We used this linear interpolation when comparing the theoretical calculations with the experimental data.

The relative crystal growth rate $u[T(x)]/u[T(0)]$ in the CL and WF formulations is shown in Fig.~\ref{X_and_u}b as blue and green solid line, respectively, with both models giving a reasonable account of the experimental data. The discrepancy observed for the argon-rich mixtures, for which the experimental rates are systematically smaller than those predicted by Eq.~(\ref{u}), might arise from the slightly different physical meaning of $k$ and $u(T)$, the former also implicitly including the contribution from the nucleation rate~\cite{Jackson2004}. Therefore, the experimental and theoretical relative rates in Fig.~\ref{X_and_u}b would be directly comparable only if the nucleation rate were independent on composition; our results indicate that for mixtures of argon and krypton this may in fact be not the case.

The comparison in Fig.~\ref{X_and_u}b between the experimental data and theoretical calculations does not allow assessing which of the two crystal growth models effectively describes the kinetics of crystallization. This can be understood by the fact that at the temperatures at which crystallization occurred in the liquid jets the crystal growth rate is largely determined by the thermodynamic factor in the square brackets of Eq.~(\ref{u}), and only at lower temperatures does the kinetic contribution from the crystal addition rate become dominant. In order to discern between the two models, we performed MD simulations of a seeded fcc crystal growth~\cite{Tang2013,Broughton1982} in liquid mixtures of argon and krypton at much lower temperatures than those attained experimentally. The simulation details are provided in Methods and Supplementary Fig.~3. Our systems crystallized via the stacking of one slab of atoms (a {\it layer}) on top of each other along either the (100) or (111) surface. We analyzed the atomic configurations by the averaged local bond order parameters to characterize the local order in the system and to evaluate the crystal growth rate (see Methods). The crystal growth rates of the (100) and (111) surfaces obtained from simulations carried out at the interpolated temperature $T(x)$ are shown in Fig.~\ref{X_and_u}b as ratios to the respective rates for pure argon of 13.3~m~s$^{-1}$ and 8.0~m~s$^{-1}$, respectively. The simulation results agree reasonably well with the theoretical calculations, evidencing no significant differences between the crystal growth rates of the two surfaces.

\begin{figure}[h]
\includegraphics[width=0.55\linewidth]{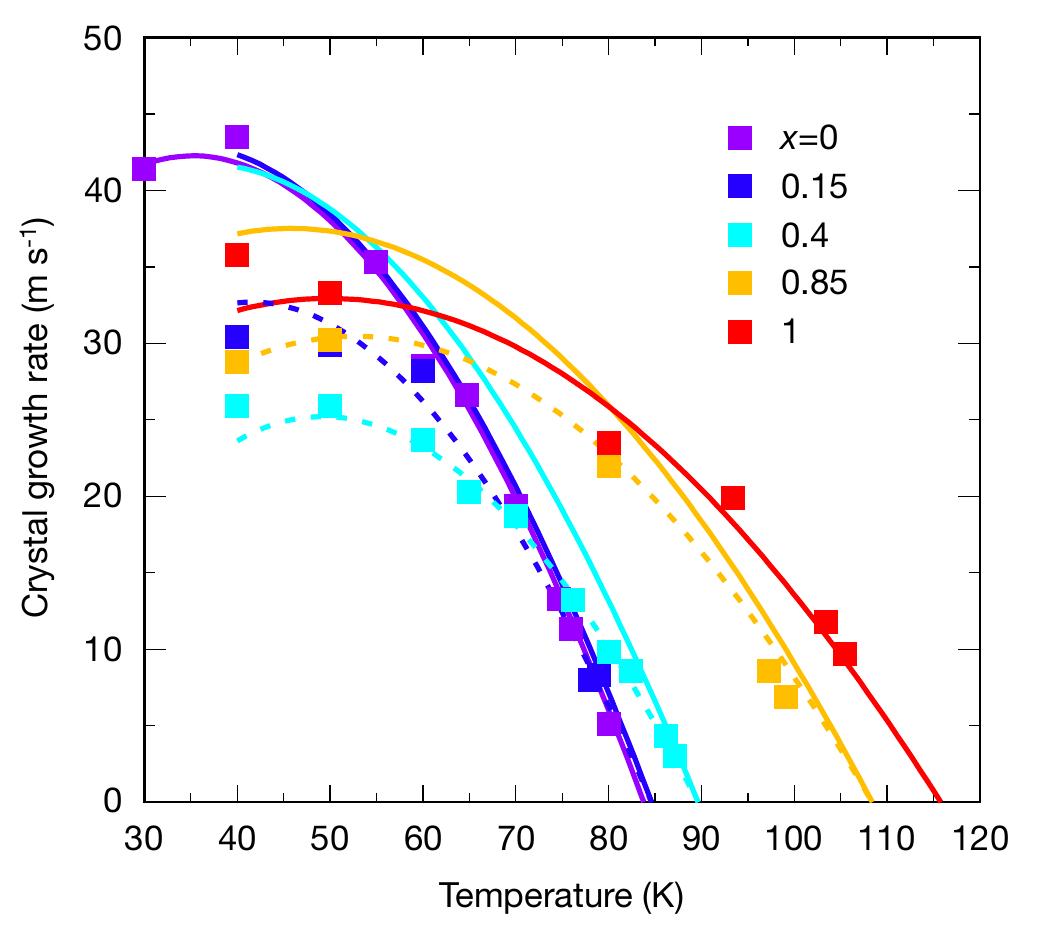}
\caption{\label{u_of_T} {\bf Temperature dependence of simulated and theoretical crystal growth rates.} The symbols are the simulation results for the crystal growth rate of the (100) surface. The error bars are smaller than the symbol size. The solid lines are calculations with the CL model, whereas the dotted lines are calculations with the modified CL crystal addition rate of Eq.~(\ref{nu_mod}). All theoretical curves were calculated with $f/c=1.9$, as obtained by fitting the CL model to the simulated crystal growth rates for pure argon.
}
\end{figure}

Having established that at the experimental jet temperatures both theory and simulations do yield consistent results, we now turn to the analysis of the temperature dependence of the simulated crystal growth rates of the surface (100), plotted in Fig.~\ref{u_of_T} for five representative systems. We determined the free parameters $f/c^2$ and $f/c$ of the WF and CL models, respectively, by fitting $u[T(x)]$ in either case to the simulated crystal growth rates (see Supplementary Fig.~4), obtaining $f/c^2=31.4$ and $f/c=1.8$. In Supplementary Fig.~5 we show that, as already established by Broughton {\it et al.} for a pure LJ liquid~\cite{Broughton1982}, the strongly activated nature of the diffusive kinetics completely fails to describe the observed temperature dependence in all simulated systems, with the WF model deviating from the simulation results already at small supercooling. By contrast, the violet and red solid lines in Fig.~\ref{u_of_T} show that the CL model provides a good description of the simulated crystal growth rates of the pure systems in the full temperature range~\cite{Broughton1982}. In these calculations we used the slightly different value $f/c=1.9$, as obtained independently by a direct fit of the CL model to the temperature dependence of the simulated crystal growth rates of pure argon. The comparison in Fig.~\ref{u_of_T}, however, also shows that the CL model is unable to account for the peak growth rates in the supercooled mixtures. In particular, the maximum theoretical crystal growth rate predicted for the $x=0.4$ mixture (cyan solid line) is roughly 80\% larger than that found in the simulations. These results provide clear evidence that neither the WF nor the CL model can properly describe the crystal growth kinetics in the simplest supercooled atomic liquid mixtures of the present study.

In the task to account for this failure of the theory, we first recall that in the liquid mixture a force $\mathbf{F}_{\alpha}=-\nabla\mu^{\rm L}_{\alpha}$ on an atom of species $\alpha\in \left\{\rm{Ar, Kr}\right\}$ is generated by the gradient of its chemical potential $\mu^{\rm L}_{\alpha}=\mu^{\rm L}_{0\alpha}+k_{B}T\ln(x_{\alpha}\gamma^{\rm L}_{\alpha})$~\cite{Taylor1993}, where $\mu^{\rm L}_{0\alpha}$ is the potential of pure species $\alpha$ at the same thermodynamic conditions of the mixture, $x_{\alpha}$ is the mole fraction, and $\gamma^{\rm L}_{\alpha}$ is the activity coefficient~\cite{Rowlinson1982}. An elementary calculation yields $\nabla\mu^{\rm L}_{\alpha}=k_{B}T\Phi\nabla\ln x_{\alpha}$, where
\begin{equation}
\Phi=1+x_{\alpha}\frac{\partial\ln\gamma^{\rm L}_{\alpha}}{\partial x_{\alpha}}
\label{Phi}
\end{equation}
is a thermodynamic factor independent on the choice of the species $\alpha$~\cite{Taylor1993}. This factor was shown to provide a correction term to the diffusion coefficient for a binary mixture~\cite{Darken1948} (see Methods), thereby accounting for effects of non-ideality of the mixture on the diffusive motion of the particles. Since in a pure LJ liquid the crystal growth kinetics as described by the CL model is determined by the short-time thermal motion characteristic of an ideal gas~\cite{Hawken2019}, the question arises how the particle's velocity is affected by $\mathbf{F}_{\alpha}$ in the mixture. An approximate analysis shows that the short-time solution to the equation of motion is represented by the average thermal velocity scaled by $\Phi$. Hence, we can define the modified CL crystal addition rate
\begin{equation}
\widetilde{\nu}_{\rm CL}(T)=\nu_{\rm CL}(T)\Phi,
\label{nu_mod}
\end{equation}
where the explicit expression of $\Phi$ is derived in Methods, see Eq.~(\ref{Phi_M}). Note that $\widetilde{\nu}_{\rm CL}(T)= \nu_{\rm CL}(T)$ for the pure systems, by definition. The crystal growth rates calculated using $\widetilde{\nu}_{\rm CL}(T)$ are shown in Fig.~\ref{u_of_T} as dotted lines. The agreement with the simulation results is now remarkable, indicating that the modified CL model successfully captured the full temperature dependence of the crystal growth rate in the supercooled mixtures. The dotted line in Fig.~\ref{X_and_u}b represents the theoretical calculation with the modified CL model at the composition-dependent temperature $T(x)$ (see inset in Fig.~\ref{exp}b), showing a slightly improved agreement with both the experimental data and simulation results for the krypton-rich mixtures.

We extended the comparison between theory and simulations also to the growth of the (111) surface. The results, presented in Supplementary Fig.~6, clearly show that the dependence on temperature of the crystal growth rate of the (111) surface is also well described by the modified CL model. In these calculations we used $f/c=1.1$, reflecting the smaller absolute crystal growth rates found in the simulations when compared to the (100) surface. It is significant that our results are in sharp contrast to those reported by Burke {\it et al.}~\cite{Burke1988}, who found that the crystal growth rate of the (111) surface in a pure supercooled LJ liquid was described by the WF model.

In conclusion, we have shown that the departure from ideality provides a simple, clear physical account of the crystal growth rates in supercooled mixtures of argon and krypton, thereby significantly improving the canonical view of crystallization. We anticipate that the crystal addition rate in Eq.~(\ref{nu_mod}) might be especially relevant to the description of crystal growth in regular solutions. As a further important example of such a binary system, the explicit calculation of $\Phi$ for alloys of copper and nickel based on available assessed data~\cite{Turchanin2007}, extrapolated to supercooled temperatures, does indicate that the modified CL model can consistently explain experimental~\cite{Willnecker1989,Algoso2003} and simulation~\cite{Fang2013} results on crystal growth in these liquids. Having established the key role played by the thermodynamic complexity that distinguishes binary liquid mixtures in the kinetics of crystal growth, the opportunity now exists for a quantitative description of crystal growth in binary systems beyond the simplest atomic liquids discussed in the present paper, and particularly in strongly non-ideal alloys exhibiting several intermediate solid phases and extended eutectic regions.

\newpage
\noindent {\bf Figure Legends}
\clearpage

\noindent{\bf METHODS}

\noindent{\bf Experimental details.} The x-ray scattering experiments were performed at the P03 beamline of the PETRA III synchrotron at DESY, Hamburg. The liquid jets were generated by gas condensation in a glass capillary cryogenically cooled to temperatures slightly above the equilibrium liquidus line of mixtures of argon and krypton~\cite{Heastie1959}. The mixtures were prepared at room temperature by continuous mixing of 99.999\% purity argon and krypton gases at specific ratios set by two mass flow controllers, one for each gas. The liquid jet velocities were determined from the Bernoulli equation $v=\sqrt{2P/\rho}$, where $P$ is the applied gas supply pressure and $\rho$ is the density of the liquid. The source pressure varied between 20~bar and 40~bar, depending on the specific mixture, providing jet velocities between 55~m s$^{-1}$ and 77~m s$^{-1}$. We verified that the calculated velocities were consistent with those determined by mass conservation from the pressure rise in the experimental vacuum chamber.

Diffraction images were recored by employing a Pilatus 300k (Dectris) pixel array detector, placed slightly sideways with respect to the x-ray beam direction at a distance of $229\pm 0.9$~mm from the jet. For analysis, the two-dimensional diffraction patterns were azimuthally integrated and background-subtracted, and subsequently corrected for polarization and geometric effects. The $q$-values were calibrated by recording diffraction patterns from the thin layer of hexagonal ice that formed by condensation at the capillary tip of water molecules invariably present as residual gas in the vacuum chamber. The background images were recorded at each $z$ at a radial distance of $\approx 50$ $\mu$m from the jet axis where no scattering from the filament was expected. However, the fact that the transmitted x-ray beam was attenuated by the liquid jet resulted in different contributions to the diffraction intensity of the x rays scattered from the thin Kapton foil placed between the jet and the detector. We removed this distortion by comparing background diffraction images measured at different distances from the nozzle. 

To analyze the contributions to the diffraction intensity of the disordered liquid and crystalline phases we used up to eight Voigt functions to fit the experimental diffraction profiles. The contribution of the disordered liquid was modeled by assuming three Voigt functions with fixed area ratios and relative peak positions. The fraction crystallized was calculated by dividing the sum of the integrated fcc peak intensities by the total integrated intensity. To estimate the uncertainty, we considered two different integration ranges, one containing the full measured diffraction profile, and one limited to $1$~\AA$^{-1}$~$\leq q \leq$~$2.5$~\AA$^{-1}$ and containing only the two most intense fcc peaks.

\noindent{\bf Jet temperature determination.} The jet temperatures at the onset of crystallization were determined from the analysis of the first visible (111) fcc peak in the integrated diffraction patterns. The associated $q$-value was used to extract, by means of Bragg's law, the lattice constant $a_{\rm exp}$ of the growing crystal in each investigated system. For a cubic unit cell, the lattice constant $a$ is related to the density $\rho$ by $a^3=NM_{\rm mol}/(N_{\rm A}\rho)$, where $M_{\rm mol}$ is the molar mass, and $N_{\rm A}$ is the Avogadro number. For the fcc crystal structure the number of atoms per unit cell is $N=4$. Accordingly, the lattice constant as a function of $x$ and $T$ becomes
\begin{equation}
a(x,T)=\left[4\frac{m(x)}{\rho(x,T)}\right]^{1/3},
\label{a}
\end{equation}
where $m(x)=(1-x)m_{\rm Ar}+xm_{\rm Kr}$ and $\rho(x,T)=(1-x)\rho_{\rm Ar}(T)+x\rho_{\rm Kr}(T)$, with $m_{\rm Ar}$ and $m_{\rm Kr}$ the masses of argon and krypton, respectively, and where the densities of solid argon and krypton $\rho_{\rm Ar}(T)$ and $\rho_{\rm Kr}(T)$, respectively, were determined from fits to experimental molar volume data~\cite{Ferreira2008}. We verified the validity of Eq.~(\ref{a}) by comparing it with lattice constants measured in solid mixtures of argon and krypton at 7 K~\cite{Kovalenko1972}. The jet temperature was then obtained as graphical solution to $a(x,T)-a_{\rm exp}=0$ for each $x$, as shown in Supplementary Fig.~2.

\noindent{\bf Theoretical crystal growth rate calculations.} We calculated the explicit analytic dependence on $x$ and $T$ of each of the quantities appearing in Eq. (1) in the main text. The expression for the lattice constant was already derived above, Eq.~(\ref{a}). For the calculations with the CL model we used the weighted mass $m(x)$ defined above. The diffusional behavior in binary mixtures was investigated experimentally by Vignes~\cite{Vignes1966}, who proposed an empirical expression for the composition dependence of the diffusion coefficient that was later justified on a theoretical ground by Cullinan~\cite{Cullinan1966}. In this formulation, the binary diffusion coefficient used in the WF model reads
\begin{equation}
D(x,T)=\left[D_{\rm Ar}(T)\right]^{1-x}\left[D_{\rm Kr}(T)\right]^{x}\Phi(x,T),
\end{equation}
where $\Phi(x,T)$ is defined by Eq.~(\ref{Phi}) in the main text. The diffusion coefficients of the pure substances were obtained as zero-pressure linear extrapolations of experimental diffusivity data at higher pressures~\cite{Naghizadeh1962} by assuming a temperature dependence of the usual Arrhenius form, $D_0\exp\left(-Q/T\right)$, with $D_0=1.21\times 10^{-7}$~m$^2$ s$^{-1}$ and $Q=350.77$~K for argon, and $D_0=0.51\times 10^{-7}$~m$^2$ s$^{-1}$ and $Q=402.51$~K for krypton.

The Gibbs free energies of the liquid and crystal are~\cite{Rowlinson1982},
\begin{equation}
G^{\rm L,C}(x,T)=(1-x)g_{\rm Ar}^{\rm L,C}(T)+xg_{\rm Kr}^{\rm L,C}(T)+g_{\rm mix}(x,T)+g_{\rm E}^{\rm L,C}(x,T),
\label{G}
\end{equation}
where $g^{\rm L,C}_{\rm Ar}(T)$ and $g^{\rm L,C}_{\rm Kr}(T)$ are the Gibbs free energies for pure argon and krypton, respectively, $g_{\rm mix}(x,T)=\left[(1-x)\ln(1-x)+x\ln x\right]RT$ is the ideal free energy of mixing, and
\begin{equation}
g_{\rm E}^{\rm L,C}(x,T)=\left[(1-x)\ln\gamma^{\rm L,C}_{\rm Ar}+x\ln\gamma^{\rm L,C}_{\rm Kr}\right]RT
\label{g_excess}
\end{equation}
is the excess free energy expressed as a function of the activity coefficients of argon and krypton~\cite{Rowlinson1982}.

The free energies of the pure substances in Eq.~(\ref{G}) were obtained from their respective heat capacities $c_p^{\rm L,C}(T)$, since $g^{\rm L,C}(T)=h^{\rm L,C}(T)-Ts^{\rm L,C}(T)$, with the enthalpies and entropies given by
\begin{equation}
h^{\rm L,C}(T)=h^{\rm L,C}(T_{\rm m})-\int_{T}^{T_{\rm m}} dT^{\prime}c_p^{\rm L,C}(T^{\prime})
\end{equation}
and
\begin{equation}
s^{\rm L,C}(T)=s^{\rm L,C}(T_{\rm m})-\int_{T}^{T_{\rm m}} dT^{\prime}\frac{c_p^{\rm L,C}(T^{\prime})}{T^{\prime}},
\end{equation}
respectively, being $T_{\rm m}$ the melting temperature. The enthalpy of fusion $\Delta h_{\rm m}=h^{\rm L}(T_{\rm m})-h^{\rm C}(T_{\rm m})$ is $\Delta h_{\rm m}=1180$ J mol$^{-1}$ for argon and $\Delta h_{\rm m}=1640$ J mol$^{-1}$ for krypton. We assumed the heat capacities of the supercooled liquids as extrapolations for $T<T_{\rm m}$ of experimental data at higher temperatures~\cite{Flubacher1961,Gladun1970,Gladun1971}, fitted by $c_{p}^{\rm L}(T)=a_0+a_1T+a_2T\exp(a_3T)$ (see Supplementary Fig.~1). The fit parameters are $a_0=28.6384$~J~mol$^{-1}$~K$^{-1}$, $a_1=0.1744$~J~mol$^{-1}$ K$^{-2}$, $a_2=4.66\times 10^{-8}$~J~mol$^{-1}$ K$^{-2}$, and $a_3=0.111$~K$^{-1}$ for argon, and $a_0=40.5517$~J~mol$^{-1}$~K$^{-1}$, $a_1=0.0349$~J~mol$^{-1}$~K$^{-2}$, $a_2=1.7169\times 10^{-6}$~J~mol$^{-1}$~K$^{-2}$, and $a_3=0.0599$~K$^{-1}$ for krypton. The experimental heat capacities of the pure solids~\cite{Flubacher1961,Beaumont1961} were fitted for $T>10$~K by the fourth-order polynomial $c_{p}^{\rm C}(T)=b_0+b_1T+b_2T^2+b_3T^3+b_4T^4$ (see Supplementary Fig.~1), with $b_0=-10.2982$~J~mol$^{-1}$~K$^{-1}$, $b_1=1.5792$~J~mol$^{-1}$~K$^{-2}$, $b_2=-0.0252$~J~mol$^{-1}$~K$^{-3}$, $b_3=1.6155\times 10^{-4}$~J~mol$^{-1}$~K$^{-4}$, and $b_4=-9.4934\times 10^{-8}$~J~mol$^{-1}$~K$^{-5}$ for argon, and $b_0=-7.9314$~J~mol$^{-1}$~K$^{-1}$, $b_1=1.7226$~J~mol$^{-1}$~K$^{-2}$, $b_2=-0.0332$~J~mol$^{-1}$~K$^{-3}$, $b_3=2.8317\times 10^{-4}$~J~mol$^{-1}$~K$^{-4}$, and $b_4=-8.3907\times 10^{-7}$~J~mol$^{-1}$~K$^{-5}$ for krypton.

For practical purposes, the excess term of Eq.~(\ref{g_excess}) is written as a series expansion~\cite{Rowlinson1982},
\begin{equation}
g_{\rm E}^{\rm L,C}(x,T)=(1-x)x\left[\xi_0^{\rm L,C}(T)+\xi_1^{\rm L,C}(T)(1-2x)+\xi_2^{\rm L,C}(T)(1-2x)^2+\dots\right]RT,
\end{equation}
where $\xi_0^{\rm L,C}(T)$, $\xi_1^{\rm L,C}(T)$, $\xi_2^{\rm L,C}(T)$, \ldots, are empirical coefficients that depend on temperature. Davies {\it et al.}~\cite{Davies1967} measured the excess free energy for equimolar liquid mixtures of argon and krypton at two different temperatures, providing the first three coefficients in the expansion of $g_{\rm E}^{\rm L}(x,T)$. The experimental values indicate that such mixtures closely approach the limit of the so-called regular solutions~\cite{Rowlinson1982}, for which $\xi_0^{\rm L,C}(T)RT=h_0^{\rm L,C}-Ts_0^{\rm L,C}$, and $\xi_1^{\rm L,C}(T)=\xi_2^{\rm L,C}(T)=\ldots=0$. The data reported in Ref.~\cite{Davies1967} yield $h_0^{\rm L}=279.07$~J~mol$^{-1}$ and $s_0^{\rm L}=-0.49$~J~mol$^{-1}$~K$^{-1}$. Similarly, Fender and Halsey investigated solid solutions of argon and krypton, and, in the regular solution approximation, they found $h_0^{\rm C}=1163.15$~J~mol$^{-1}$ and $s_0^{\rm C}=5.48$~J~mol$^{-1}$~K$^{-1}$~\cite{Fender1965}.

The entropy of fusion in Eq.~(1) in the main text is a function of the mole fraction only, $\Delta S_{\rm m}(x)=\left\{H^{\rm L}\left[x,T_{\rm m}(x)\right]-H^{\rm C}\left[x,T_{\rm m}(x)\right]\right\}/T_{\rm m}(x)$, with the enthalpies $H^{\rm L,C}(x,T)=G^{\rm L,C}(x,T)-T\partial G^{\rm L,C}(x,T)/\partial T$ evaluated at the liquidus temperature $T_{\rm m}(x)$. We obtained an analytic expression for $T_{\rm m}(x)$ by fitting a sixth-order polynomial to the experimental phase-equilibrium data~\cite{Heastie1959}, as shown in the inset of Fig.~\ref{exp}b in the main text.

To evaluate the thermodynamic factor of Eq.~(\ref{Phi}) in the main text in terms of the experimental excess data for mixtures of argon and krypton, we first write the excess free energy for the liquid as
\begin{equation}
g_{\rm E}^{\rm L}(x,T)=(1-x)x\xi_0^{\rm L}(T)RT=\left[(1-x)x^2\xi_0^{\rm L}(T)+x(1-x)^2\xi_0^{\rm L}(T)\right]RT.
\end{equation}
By comparing the last expression in the above equation with Eq.~(\ref{g_excess}), we find for example $\ln\gamma^{\rm L}_{\rm Kr}=(1-x)^2\xi_0^{\rm L}(T)$, and thus
\begin{equation}
\Phi(x,T)=1+x\frac{\partial\ln\gamma^{\rm L}_{\rm Kr}}{\partial x}=1-2(1-x)x\xi_0^{\rm L}(T).
\label{Phi_M}
\end{equation}

\noindent{\bf Simulation details.}
We performed the MD simulations using LAMMPS (\url{http://lammps.sandia.gov}) on CINECA high-performance computing facilities (\url{https://www.cineca.it}). The simulation box, shown in Supplementary Fig.~3, was a rectangular parallelepiped with its longer side oriented along the crystal growth direction. Periodic boundary conditions were applied along all three spatial directions. At each composition, we investigated the growth of the fcc (100) and (111) surfaces. The simulation box for the (100) crystal growth contained 9216 atoms ($8\times8\times36$ unit cells), whereas that for the (111) crystal growth contained 8640 atoms ($10\times6\times36$ unit cells). To rule out size effects, we performed simulations also with larger systems containing 20736 atoms, finding consistent results. We employed the standard velocity Verlet integration scheme with a time step of 2~fs. The atoms interacted through a LJ potential $v_{\alpha \beta}(r)=4\epsilon_{\alpha \beta}\left[\left(\sigma_{\alpha \beta}/r\right)^{12}-\left(\sigma_{\alpha \beta}/r\right)^6\right]$, $\alpha,\beta\in \left\{\rm{Ar, Kr}\right\}$, whose parameters for the pure systems were $\epsilon_{\rm ArAr}=0.238$~kcal~mol$^{-1}$, $\sigma_{\rm ArAr}=3.4$~\AA, $\epsilon_{\rm KrKr}=0.318$~kcal~mol$^{-1}$, and $\sigma_{\rm KrKr}=3.64$~\AA. The parameters for the potential between distinct species obeyed the Lorentz-Berthelot relations $\epsilon_{\rm ArKr}=\sqrt{\epsilon_{\rm ArAr}\epsilon_{\rm KrKr}}$ and $\sigma_{\rm ArKr}=\left(\sigma_{\rm ArAr}+\sigma_{\rm KrKr}\right)/2$. The potential cut-off was set at 20.4~\AA\ for all simulated systems. The atomic masses were $m_{\rm Ar}=39.948$~amu and $m_{\rm Kr}=83.798$~amu.

The simulation box accommodated two independent crystal growths in opposite directions with respect to a central crystalline seed of either pure argon -- for argon-rich systems -- or pure krypton -- for krypton-rich systems. The particles of the seed interacted via the standard LJ potential, but they were additionally confined to their equilibrium positions by a single-particle harmonic potential whose elastic constant was determined to reproduce the average atomic positions of a reference crystalline fcc phase at fixed temperature. The region that we used to perform the statistical analysis comprised eighteen atomic layers labeled by the index $l=1,\dots , 18$ in each half of the simulation box. The analysis region was separated from the central seed by a buffer region consisting of four additional layers. The atoms in the analysis region were time-integrated at constant atom number, volume, and energy. Both the crystalline seed and the buffer were kept at constant temperature by a Bussi-Parrinello thermostat. As crystallization proceeded, the thermostat was progressively extended to the as-solidified layers at a pace slightly lower than the growth rate in order to dissipate the latent heat. A supercooled liquid region separated from the analysis region by additional buffer layers was used to keep the box at zero pressure via a Berendsen barostat and as a particle reservoir to maintain a constant chemical potential~\cite{Tang2013}, thereby guaranteeing a uniform crystal growth. The barostat acted only along the growth direction to leave the transverse crystal lattice constant unchanged. Each simulation started by equilibrating the whole system above the melting temperature for $8 \times 10^3$~fs, during which the central seed retained its crystal structure because of the harmonic confinement. The system was then cooled within 100~fs to the final temperature. With the input parameters reported in Supplementary Table~2, the time for crystallization of the supercooled liquid varied between $\approx 3.6\times 10^5$~fs for pure argon and $\approx 2.2\times 10^6$~fs for the $x=0.4$ mixture.

The MD trajectories were analyzed by means of the structural order parameter
\begin{equation}
S(i) = \sum_{j=1}^{N(i)} \frac{ \textbf{q}_6(i) \cdot \textbf{q}^*_6(j)}{|\textbf{q}_6(i)|\:|\textbf{q}_6(j)|}\, ,
\end{equation}
where $N(i)$ is the number of first neighbors of the $i$th particle and where the 13-dimensional vector $\textbf{q}_6(i)$ is one of the standard rotationally invariant Steinhardt parameters that allow distinguishing between different crystal symmetries~\cite{Steinhardt1983}. Typically, $S(i)$ ranges from $\approx 0.2$ for a disordered liquid to $\approx 0.9$ for a crystal. To determine the crystal growth rate, we labeled a layer with index $l$ as the {\it crystal front} (CF) if the crystallinity condition $S(i) \geq 0.7$ was satisfied by at least 50\% of the atoms in the layers with indices $l$ and $l-1$. Indicating with $\tau^{\rm CF}_l$ the time at which the layer of index $l$ became the crystal front, we computed the crystal growth rate according to
\begin{equation}
u_{l^{\prime}}=\frac{L}{\tau^{\rm CF}_{l^{\prime}+9} - \tau^{\rm CF}_{l^{\prime}}}\, ,
\end{equation}
with $l^{\prime}=1,\dots , 9$, and where $L$ is half the length of the analysis region (see Supplementary Fig.~3). In this way, we obtained a total of eighteen independent estimations of the crystal growth rate for each simulation run.

\noindent{\bf Code availability.} 
The codes used during the current study are available from the corresponding author upon reasonable request.
\\

\noindent{\bf ACKNOWLEDGMENTS}

\noindent We acknowledge financial support from the Bundesministerium f\"ur Bildung und Forschung (Grant No. 05K13RF5). We acknowledge the CINECA awards LISA-PUMAS (2016), IscraC-GLEMD (2017) and IscraB-MEMETICO (2018) for the availability of high performance computing resources and support. We acknowledge DESY (Hamburg, Germany), a member of the Helmholtz Association HGF, for the provision of experimental facilities. Parts of this research were carried out at PETRA III and we would like to thank Stephan Roth for the unlimited support during the experiments at the beamline P03.
\\

\noindent{\bf AUTHOR CONTRIBUTIONS}

\noindent A. S. and F. M. contributed equally to this work. R. E. G. conceived the experiment. A. S. designed and assembled the experimental setup. A. S., F. M., A. K., B. B., A. R., C. G., J. M., N. P., M. R., F. T., J. M. F., T. A. E., and R. E. G. performed the experiments. A. S. and R. E. G. analyzed the experimental data. R. E. G. performed the theoretical crystal growth rate calculations. F. M. and D. E. G. conceived and carried out the MD simulations. F. M., D. E. G., and R.~E.~G. wrote the paper, with contributions from all authors
\\

\noindent{\bf COMPETING FINANCIAL INTERESTS}

\noindent The authors declare no competing financial interests.
\\

\noindent{\bf DATA AVAILABILITY}

\noindent The datasets generated and/or analyzed during the current study are available from the corresponding author on reasonable request.



\begin{thebibliography}{00}

\bibitem{Tang2013}
C. Tang and P. Harrowell. Anomalously slow crystal growth of the glass-forming alloy CuZr. Nat. Mater. \textbf{12}, 507 (2013).

\bibitem{Orava2014}
J. Orava and A. L. Greer. Fast and slow crystal growth kinetics in glass-forming melts. J. Chem. Phys. \textbf{140}, 214504 (2014).

\bibitem{Orava2016}
J. Orava and A. L. Greer. Fast crystal growth in glass-forming liquids. J. Non-Cryst. Solids \textbf{451}, 94 (2016).

\bibitem{Kelton2010}
K. F. Kelton and A. L. Greer. Nucleation in Condensed Matter (Elsevier, Amsterdam, 2010).

\bibitem{Jackson2004}
K. A. Jackson. Kinetic Processes (Wiley, Weinheim, 2004).

\bibitem{Spaepen1984}
F. Spaepen and C. J. Lin. Partitionless crystallization and glass formation in Fe-B alloys during picosecond pulsed laser quenching. In Amorphous Metals and Non-Equilibrium Processing, ed. M. von Allmen (Les Ulis: Les Editions de Physique, France, 1984).

\bibitem{Kerrache2008}
A. Kerrache, J. Horbach, and K. Binder. Molecular-dynamics computer simulation of crystal growth and melting in Al$_{50}$Ni$_{50}$. Europhys. Lett. \textbf{81}, 58001 (2008).

\bibitem{Stipp2009}
A. Stipp and T. Palberg. Crystal growth kinetics in binary mixtures of model charged sphere colloids. Phil. Mag. Lett. \textbf{87}, 899 (2007).

\bibitem{Wang2011}
Q. Wang, Li-Min Wang, M. Z. Ma, S. Binder, T. Volkmann, D. M. Herlach, J. S. Wang, Q. G. Xue, Y. J. Tian, and R. P. Liu. Diffusion-controlled crystal growth in deeply undercooled Zr$_{50}$Cu$_{50}$ melt on approaching the glass transition. Phys. Rev. B \textbf{83}, 014202 (2011).

\bibitem{Fang2013}
T. Fang, L. Wang, and Y. Qi. Solid-liquid interface growth of Cu$_{50}$Ni$_{50}$ under deep undercoolings. Phys. Chem. Liquids \textbf{52}, 342 (2013).

\bibitem{Yan2015}
X. Q. Yan and Y. J. L\"u. Mechanism of abnormally slow crystal growth of CuZr alloy. J. Chem. Phys. \textbf{143}, 164503 (2015).

\bibitem{Sun2019}
Y. Sun, F. Zhang, L. Yang, H. Song, M. I. Mendelev, C.-Z. Wang, and K.-M. Ho. Effects of dopants on the glass forming ability in Al-based metallic alloy. Phys. Rev. Mater. \textbf{3}, 023404 (2019).

\bibitem{Frenkel1946}
Y. Frenkel. The Kinetic Theory of Liquids (Oxford Univ. Press, 1946).

\bibitem{Broughton1982}
J. Q. Broughton, G. H. Gilmer, and K. A. Jackson. Crystallization Rates of a Lennard-Jones Liquid. Phys. Rev. Lett., \textbf{49}, 1496 (1982).

\bibitem{Hawken2019}
A. Hawken, G. Sun, and P. Harrowell. Role of interfacial inherent structures in the fast crystal growth from molten salts and metal. Phys. Rev. Mater. \textbf{3}, 043401 (2019).

\bibitem{Sun2018}
G. Sun, J. Xu, and P. Harrowell. The mechanism of the ultrafast crystal growth of pure metals from their melts. Nat. Mater. \textbf{17}, 881 (2018).

\bibitem{Burke1988}
E. Burke, J. Q. Broughton, and G. H. Gilmer. Crystallization of FCC (111) and (100) crystal-melt interfaces: a comparison by molecular dynamics for the Lennard-Jones system. J. Chem. Phys. \textbf{89}, 1030 (1988).

\bibitem{Heastie1959}
R. Heastie. Properties of solid and liquid solutions of argon and krypton. Proc. Phys. Soc. \textbf{73}, 490 (1959).

\bibitem{Kuehnel2011}
M. K\"uhnel, J. M. Fern\'andez, G. Tejeda, A. Kalinin, S. Montero, and R. E. Grisenti. Time-resolved study of crystallization in deeply cooled liquid parahydrogen. Phys. Rev. Lett. \textbf{106}, 245301 (2011).

\bibitem{Grisenti2018}
R. E. Grisenti, A. Kalinin, C. Goy, and A. Schottelius. Evaporating laminar microjets for studies of rapidly evolving structural transformations in supercooled liquids. Adv. Phys. X \textbf{3}, 1418183 (2018).

\bibitem{Pollack1964}
G. L. Pollack. The solid state of rare gases. Rev. Mod. Phys. \textbf{36}, 748 (1964).

\bibitem{Cahn1956}
J. W. Cahn. Transformation kinetics during continuous cooling. Acta Metallurgica \textbf{4}, 572 (1956).

\bibitem{Williams2008}
S. R. Williams, C. P. Royall, and G. Bryant. Crystallization of dense binary hard-sphere mixtures with marginal size ratio. Phys. Rev. Lett. \textbf{100}, 225502 (2008).

\bibitem{Kuehnel2014}
M. K\"uhnel, J. M. Fern\'andez, F. Tramonto, G. Tejeda, E. Moreno, A. Kalinin, S. Montero, M. Nava, D. E. Galli, and R. E. Grisenti. Observation of crystallization slowdown in supercooled parahydrogen and orthodeuterium quantum liquid mixtures. Phys. Rev. B \textbf{89}, 180201(R) (2014).

\bibitem{Taylor1993}
R. Taylor and R. Krishna. Multicomponent Mass Transfer (Wiley, New York, 1993).

\bibitem{Rowlinson1982}
J. S. Rowlinson and F. L. Swinton. Liquids and Liquid Mixtures (Butterworth \& Co, London, 1982).

\bibitem{Darken1948}
L. S. Darken. Diffusion, mobility and their interrelation through free energy in binary metallic systems. Trans. Am. Inst. Min. Metall. Eng. \textbf{175}, 184 (1948).

\bibitem{Turchanin2007}
M. A. Turchanin, P. G. Agraval, and A. R. Abdulov. Phase equilibria and thermodynamics of binary copper systems with 3d-metals. VI. Copper-Nickel system. Powder Metall. Met. Ceram. {\bf 45}, 143 (2007).

\bibitem{Willnecker1989}
R. Willnecker, D. M. Herlach, and B. Feuerbacher. Evidence of nonequilibrium processes in rapid solidification of undercooled metals. Phys. Rev. Lett. \textbf{62}, 2707 (1989).

\bibitem{Algoso2003}
P. R. Algoso, W. H. Hofmeister, and R. J. Bayuzick. Solidification velocity of undercooled Ni-Cu alloys. Acta Mater. {\bf 51}, 4307 (2003).

\end{thebibliography}

\begin{thebibliography}{99}
\makeatletter
\addtocounter{NAT@ctr}{30}
\makeatother

\bibitem{Ferreira2008}
A. G. M. Ferreira and L. Q. Lobo. The sublimation of argon, krypton, and xenon. J. Chem. Thermodynamics \textbf{40}, 1621 (2008).

\bibitem{Kovalenko1972}
S. I. Kovalenko, E. I. Indan, and A. A. Khudoteplaya. An electron-diffraction study of thin films of the binary mixtures of rare gases. Phys. Stat. Sol. {\bf 13}, 235 (1972).

\bibitem{Vignes1966}
A. Vignes. Diffusion in binary solutions. Ind. Eng. Chem. Fundamentals {\bf 5}, 189 (1966).

\bibitem{Cullinan1966}
H. T. Cullinan, Jr. Concentration dependence of the binary diffusion coefficient. Ind. Eng. Chem. Fundamentals {\bf 5}, 281 (1966).

\bibitem{Naghizadeh1962}
J. Naghizadeh and S. A. Rice. Kinetic theory of dense fluids. X. Measurement and interpretation of self-diffusion in liquid Ar, Kr, Xe, and CH$_4$. J. Chem. Phys. \textbf{36}, 2710 (1962).

\bibitem{Flubacher1961}
P. Flubacher, A. J. Leadbetter, and J. A. Morrison. A low temperature adiabatic calorimeter for condensed substances. Thermodynamic properties of argon. Proc. Phys. Soc. \textbf{78}, 1449 (1961).

\bibitem{Gladun1970}
C. Gladun and F. Menzel. The specific heat and some other thermodynamic properties of liquid krypton. Cryogenics \textbf{10}, 210 (1970).

\bibitem{Gladun1971}
C. Gladun. The specific heat of liquid argon. Cryogenics \textbf{11}, 205 (1971).

\bibitem{Beaumont1961}
R. H. Beaumont, H. Chihara, and J. A. Morrison. Thermodynamic Properties of Krypton. Vibrational and Other Properties of Solid Argon and Solid Krypton. Proc. Phys. Soc. \textbf{78}, 1462 (1961).

\bibitem{Davies1967}
R. H. Davies, A. G. Duncan, G. Saville, and L. A. K. Staveley. Thermodynamics of liquid mixtures of argon and krypton. Trans. Faraday Soc. \textbf{63}, 855 (1967).

\bibitem{Fender1965}
B. E. F. Fender and G. D. Halsey, Jr. Solid solution of argon and krypton; Refined measurements. J. Chem. Phys. \textbf{42}, 127 (1965).

\bibitem{Steinhardt1983}
P. Steinhardt, D. R. Nelson, and M. Ronchetti. Bond-orientational order in liquids and glasses. Phys. Rev. B \textbf{28}, 784 (1983).

\end{thebibliography}
\end{document}